\documentclass[twocolumn,letter]{jpsj3}
\usepackage{txfonts}
\usepackage{color}
\usepackage{ulem}

\title{
Superconductivity in BaPtSb with an Ordered Honeycomb Network
}

\author{
Kazutaka Kudo$^1$\thanks{kudo@science.okayama-u.ac.jp}, Yuki Saito$^1$, Takaaki Takeuchi$^1$, Shin-ya Ayukawa$^1$, Takayuki Kawamata$^2$, 
\\ Shinichiro Nakamura$^2$, Yoji Koike$^2$, and Minoru Nohara$^1$\thanks{nohara@science.okayama-u.ac.jp}
}

\inst{
$^1$Research Institute for Interdisciplinary Science, Okayama University, Okayama 700-8530, Japan
\\ 
$^2$Department of Applied Physics, Tohoku University, Sendai 980-8579, Japan
}

\abst{
Superconductivity in BaPtSb with the SrPtSb-type structure (space group $P\bar{6}m2$, $D_{3h}^1$, No. 187) is reported. 
The structure consists of a PtSb ordered honeycomb network that stacks along the $c$-axis so that spatial inversion symmetry is broken globally. 
Electrical resistivity and specific-heat measurements revealed that the compound exhibited superconductivity at 1.64 K. 
The noncentrosymmetric structure and the strong spin-orbit coupling of Pt and Sb make BaPtSb an attractive compound for studying the exotic superconductivity predicted for a honeycomb network. 
}

\begin{document}
\maketitle

Superconductors with honeycomb networks have attracted considerable interest since the theoretical predictions of exotic superconductivity in SrPtAs \cite{Goryo,Fischer,Wang_Sr,Youn,Akbari,Sigrist,Fischer2,Goryo2}. 
The compound SrPtAs exhibits superconductivity at 2.4 K\cite{Nishikubo}, and 
crystallizes in the KZnAs-type structure ($P6_3/mmc$, $D_{6h}^4$, No. 194). This structure is a ternary ordered variant of the AlB$_2$-type structure. 
The Sr occupies the Al site, and the Pt and As atoms alternately occupy the B site of the honeycomb layers, thus forming a PtAs ordered honeycomb network.
The honeycomb layers are stacked along the $c$-axis in such a manner that there is an As atom above each Pt, and vice versa, over every As and Pt atom, as shown in Fig. 1(a). 
Hence, the structure is globally centrosymmetric, although the spatial inversion symmetry is locally broken in the PtAs honeycomb network. 
Possible superconducting states theoretically predicted for SrPtAs include a singlet-triplet mixed state\cite{Goryo}, a chiral $d$-wave state\cite{Fischer}, and an $f$-wave state\cite{Wang_Sr}.

The above-mentioned predictions have been examined experimentally. 
$\mu$SR measurements reported the breaking of time-reversal symmetry in the superconducting state, suggesting the chiral $d$-wave superconductivity as the most likely pairing state\cite{Biswas}, whereas the NMR/NQR\cite{Matano} and the magnetic penetration depth\cite{Landaeta} measurements suggested a conventional $s$-wave pairing state. 
Another NQR measurement showed the absence of a Hebel-Slichter coherence peak and suggested two-gap superconductivity\cite{Bruckner}. 
In order to settle this controversy, novel compounds with honeycomb networks should be developed and examined both experimentally and theoretically.

In this letter, we report the emergence of superconductivity at 1.64 K in BaPtSb with a PtSb ordered honeycomb network. 
The compound crystallizes in the SrPtSb-type structure ($P\bar{6}m2$, $D_{3h}^1$, No. 187), as reported by G. Wenski and A. Mewis in Ref. \citen{Wenski}. 
This structure is another ternary ordered variant of the AlB$_2$-type structure: As shown in Fig. 1(b), the structure is composed of alternate stacking of Ba and ordered PtSb honeycomb layers, which are stacked along the $c$-direction in such a way that there is a Pt(Sb) atom above each Pt(Sb). 
Thus, the spatial inversion symmetry is globally broken for BaPtSb, in contrast with SrPtAs, although the structure is neither polar nor chiral because of the presence of mirror symmetries. 
The noncentrosymmetric structure, the strong spin-orbit coupling of both Pt and Sb, and the honeycomb network make BaPtSb a unique compound to examine whether the exotic superconductivity predicted for honeycomb networks is realized. 
\begin{figure}[t]
\begin{center}
\includegraphics[width=8.5cm]{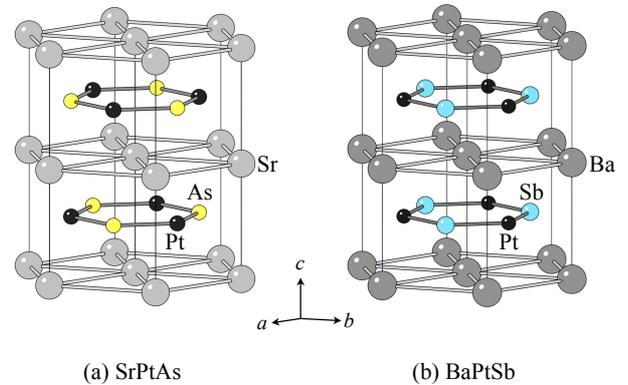}
\caption{
(Color online) Crystal structures of (a) SrPtAs with a KZnAs-type structure ($P6_3/mmc$, $D_{6h}^4$, No. 194) and (b) BaPtSb with a SrPtSb-type structure ($P\bar{6}m2$, $D_{3h}^1$, No. 187). 
}
\end{center}
\end{figure}
\begin{figure}[t]
\begin{center}
\includegraphics[width=7.5cm]{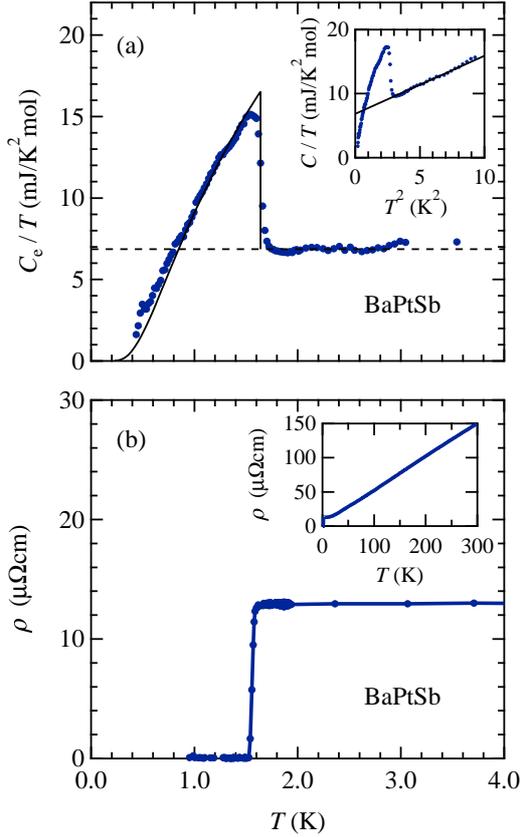}
\caption{
(Color online) Temperature dependence of (a) the electronic specific heat divided by temperature, $C_{\rm e}/T$, and (b) the electrical resistivity $\rho$ for BaPtSb. 
The solid curve shown in (a) denotes the $C_{\rm e}/T$ vs $T$ plot calculated on the basis of the weak coupling BCS approximation\cite{Muhlschlegel}. 
The inset of (a) shows the $C/T$ vs. $T^2$ plot. 
The solid line denotes the fit by $C/T = \gamma + \beta T^{2}$, where $\gamma$ is the electronic specific-heat coefficient and $\beta$ is a constant corresponding to the Debye phonon contributions. The inset of (b) shows the electrical resistivity in a wide temperature range. 
}
\end{center}
\end{figure}
%

Single crystals of BaPtSb were synthesized by heating a mixture of Ba, Pt, and Sb powders with a ratio of Ba$:$Pt$:$Sb = 1.5$:$1$:$1.2. 
The mixture was placed in an alumina crucible and sealed in a quartz tube under partial Ar pressure to minimize evaporation of Ba and Sb during the reaction. 
The powders were heated to 1150 $^{\circ}$C, and cooled to 1100 $^{\circ}$C at a rate of 2.08 $^{\circ}$C/h, followed by furnace cooling. 
The typical dimensions of the crystals were 0.2--0.5 $\times$ 0.2--0.5 $\times$ 0.2--0.5 mm$^{3}$. 
The obtained samples were characterized by powder X-ray diffraction using a Rigaku MiniFlex600 X-ray diffractometer with Cu$K_{\alpha}$, and the samples were confirmed to have a SrPtSb-type structure\cite{Wenski}. 
The specific heat $C$ was measured using the Physical Property Measurement System (PPMS) by Quantum Design. 
The measurements of electrical resistivity $\rho$ were carried out using the Adiabatic Demagnetization Refrigerator (ADR) option of the PPMS.

The emergence of bulk superconductivity was observed in the temperature dependence of the specific heat $C$ for BaPtSb. 
As shown in Fig. 2(a), the electronic specific heat $C_{\rm e}$ exhibited a clear jump, which is a hallmark of bulk superconductivity. 
The assumption of an ideal jump at the superconducting transition to satisfy the entropy conservation yielded estimates of $T_{\rm c}$ = 1.64 K and $\Delta C/T_{\rm c}$ = 9.80 mJ/K$^2$mol. 
As shown in the inset of Fig. 2(a), the normal-state data could be fitted by $C/T = \gamma + \beta T^2$, where $\gamma$ is an electronic specific heat coefficient and $\beta$ is the coefficient of phonon contribution from which the Debye temperature $\Theta_{\rm D}$ is estimated. 
The analysis yielded $\gamma$ = 6.86 mJ/K$^2$mol and $\Theta_{\rm D}$ = 186 K. 
Using this $\gamma$ value, we estimated the normalized jump $\Delta C/\gamma T_{\rm c}$ to be 1.43, which is comparable to the value expected from the BCS weak-coupling limit (1.43). 
These values can be compared with $\gamma$ = 7.31 mJ/K$^2$mol, $\Theta_{\rm D}$ = 241 K, and $\Delta C/\gamma T_{\rm c}$ = 1.07 for SrPtAs\cite{Kudo}.
The $\gamma$ value of BaPtSb was found to be comparable to that of SrPtAs, although the $T_{\rm c}$ of BaPtSb was significantly lower than that of SrPtAs. 
On the other hand, the smaller $\Theta_{\rm D}$ of BaPtSb compared with SrPtAs may be attributed to the difference in the atomic mass.

The emergence of superconductivity in BaPtSb was also evidenced by the temperature dependence of electrical resistivity, as shown in Fig. 2(b). 
The $\rho$ value decreased with decreasing temperature and exhibited residual resistivity of 13 $\mu\Omega$cm, followed by a sharp drop below 1.6 K, characteristic of a superconducting transition. 
The resistivity became negligibly small below 1.58 K, which was consistent with the value of $T_{\rm c}$ estimated from the temperature dependence of specific heat. 
The non-saturating behavior of $\rho$ at high temperatures, as shown in the inset to Fig. 2(b), suggested weak coupling between electrons and phonon.

SrAlGe and BaAlGe have been known to crystallize in the SrPtSb-type structure, which are isostructural to the present BaPtSb compound, and exhibit superconductivity at relatively high transition temperatures of 6.7 and 6.3 K, respectively\cite{Evans}. 
These compounds can be viewed as ``nine-electron'' systems, namely, nine valence electrons per formula unit. 
Among them, six electrons occupy $sp^2$ $\sigma$ bonding orbitals, which form the AlGe ordered honeycomb network, and two electrons occupy the $p_z$ $\pi$ bonding orbital. 
The remaining one electron partially occupies the $p_z$ $\pi^*$ anti-bonding orbital, which results in structural instability and therefore a relatively high superconducting transition temperature in this class of materials\cite{Evans}. 
On the other hand, the present BaPtSb compound, as well as SrPtAs, is a ``17-electron'' system, and obviously the simple picture of the $sp^2$ honeycomb network should not be valid. The one less electron, compared with a completely occupied 18-electron count, should result in partially occupied Pt 5$d$ and Sb 5$p$ orbitals, which are subjected to strong spin-orbit coupling. 
Thus, BaPtSb and SrPtAs provide a unique playground for fundamental studies of spin-orbit coupling phenomena in this class of superconductors with broken and preserved spatial inversion symmetry.

In conclusion, the specific heat and electrical resistivity measurements showed the emergence of superconductivity at 1.64 K in BaPtSb with the SrPtSb-type structure ($P\bar{6}m2$, $D_{3h}^{1}$, No. 187). 
In BaPtSb, spatial inversion symmetry is globally broken, in contrast to SrPtAs. 
The compound should be a useful platform for studying exotic superconductivity because of the noncentrosymmetric structure, the strong spin-orbit coupling of both Pt and Sb, and the honeycomb network.

\acknowledgments
A part of this work was performed at the Advanced Science Research Center, Okayama University.
This work was partially supported by Grants-in-Aid for Scientific Research (15H05886 and 16K05451) provided by the Japan Society for the Promotion of Science (JSPS) and the Program for Advancing Strategic International Networks to Accelerate the Circulation of Talented Researchers (R2705) from JSPS.

\end{document}